
\newcount\fcount \fcount=0
\def\ref#1{\global\advance\fcount by 1
\global\xdef#1{\relax\the\fcount}}

\def\simle{\lower 2pt \hbox {$\buildrel < \over {\scriptstyle \sim }$}}
\def\simge{\lower 2pt \hbox {$\buildrel > \over {\scriptstyle \sim }$}}

\def\la{\lower.5ex\hbox{$\; \buildrel < \over \sim \;$}}
\def\ga{\lower.5ex\hbox{$\; \buildrel > \over \sim \;$}}
\def\j21{J_{-21}}
\def\r{\hangindent=1pc \noindent}
\def\etal {{\it et~al.}}

\def\MeV{\,{\rm MeV}}
\def\GeV{\,{\rm GeV}}

\def\gcm2{\,{\rm g\,cm}^{-2}}

\def\erg/s{{\rm erg/s}}

\font\huge=cmr10 scaled\magstep2
\font\large=cmr10 scaled\magstep1
\font\sc=cmr10 scaled\magstep0

\magnification=\magstep1

\raggedbottom
\footline={\ifnum\pageno=1 \hfil \else \hss\tenrm\folio\hss\fi}

\tolerance = 30000

\baselineskip=5.5mm plus 0.1mm minus 0.1mm

\centerline{\bf \huge{ Gamma line radiation from stellar winds in Orion
complex:}}
\centerline{\bf \huge{ A test for cosmic ray origin theory}}
 \vskip 1.0cm
\centerline{\large{B{\sc{IMAN}} B. N{\sc{ATH}} {\sc{AND}}
P{\sc{ETER}} L. B{\sc{IERMANN}}}}
\vskip 0.2cm
\centerline{Max-Planck-Institut f\"ur Radioastronomie,
Auf dem H\"ugel 69, D-53121 Bonn, Germany}
 \centerline{\it Received \qquad\qquad\qquad; \quad accepted
  \qquad\qquad\qquad}
\vskip 1.5cm

\centerline{ABSTRACT}
\medskip
We consider $\gamma$-ray emission from heavy nuclei that
are accelerated in the shocks of stellar winds and are
excited in interactions with the ambient matter
in the Orion complex. We show that the recent
detection of $\gamma$-ray lines in the Orion complex by
COMPTEL telescope can be explained by
such a scenario, assuming cosmic abundances.
The scenario is consistent with recent
models of cosmic ray acceleration in stellar winds of
massive stars.
\bigskip

\noindent
{\it Subject headings:} ISM:  cosmic rays; gamma rays:  theory;
stars:  mass loss
\bigskip

\noindent

\vskip 1cm
\centerline{1. INTRODUCTION}
\medskip
Bloemen \etal (1994) have recently reported the detection of
$\gamma$-rays in $3-7$ MeV range in the direction of the
Orion complex with the COMPTEL telescope.
The flux observed in that energy range is
$\sim 10^{-4}$ photon cm$^{-2}$ s$^{-1}$. Subtracting
the background radiation in a $3^ \circ$ region covering Orion
A and B, they suggested that the observed photons are dominated
by lines at $4.4$ and $6.1$ MeV, and tentatively identified them
with the de-excitation lines of $^{12}$C$^\ast$ and $^{16}$O$^\ast$,
respectively (also see, Cameron 1994).

That these $\gamma$-ray lines should be emitted in interactions
between high energy particles and interstellar nuclei,
was anticipated long ago (see, for example, Meneguzzi and Reeves
1975; Ramaty, Kozlovsky and Lingenfelter 1979).
Bloemen \etal (1994) claimed that the observed line width is as large
as $\sim 1 $ MeV and inferred that the lines result from
collisions of energetic C and O nuclei with ambient protons
or $\alpha$ nuclei (rather than collisions of low energy cosmic ray protons
or $\alpha$-particles with ambient C and O nuclei in the gas and
dust). They, however, doubted that standard abundances of C and O
nuclei in cosmic rays could explain the observed luminosity.
Bykov and Bloemen (1994) have recently invoked the overabundance of
oxygen nuclei that is predicted in supernovae ejecta of massive
stars to explain the observed $\gamma$-ray photons.

In this {\it Letter} we show that cosmic ray C and O nuclei,
accelerated in the shocks in stellar
winds from massive O and B stars, can explain the observed
$\gamma$-ray luminosity, assuming cosmic abundances.
Shocks in the stellar winds of massive
stars have been previously suggested as the accelerators of cosmic
rays to moderate energies to energies as high as $ 3 \times 10^9$ GeV
(Biermann 1993; Biermann and Cassinelli 1993; Stanev, Biermann, Gaisser
1993). Such shocks have also been claimed to
generate X-rays (MacFarlane and Cassinelli 1989) and $\gamma$
rays from $\pi ^o$ decays (White 1985; White and Chen 1992).
Below we show that
the COMPTEL observations can be understood in terms of these
shocks in the stellar winds that abound in the OB associations
of the Orion complex.
\bigskip

\centerline{2. STELLAR WINDS AND $\gamma$-RAYS}
\medskip
The scenario for the production of $\gamma$-ray line photons
is briefly sketched as follows: heavy nuclei are accelerated in
the shocks in the stellar winds of OB stars and these nuclei
are excited to emit $\gamma$-ray photons when they interact
with cool gas surrounding these stars or the OB associations.
In Orion A, for example, the OB association is embedded in
a hot region which interacts with the surrounding dense and
cool molecular gas whose density is $\ga 10^5$ cm$^{-3}$
and kinetic temperature is $T \sim 100$ K
(the Orion-KL region; Genzel and
Stutzki 1989). Below, we will calculate the
$\gamma$-ray luminosity expected in this scenario.

The cross-section of the interaction between carbon nuclei
and protons that leads to the emission of a $4.4$ MeV photon
by the excited carbon nuclei has been discussed by Ramaty,
Kozlovsky and Lingenfelter 1979. They further noted that the
$4.4$ MeV level in $^{12}$C$^\ast$ can be more efficiently populated by
spallation
interaction, namely, $\sigma \lbrack ^{16}O (p, \gamma _{4.4}
 )^{12}C \rbrack \sim 10^{-25} ( {E \over 25 {\rm MeV}})^{-0.6}$
cm$^{-2}$ for $E \ga 25$ MeV (Ramaty, Kozlovsky and Lingenfelter 1979,
their fig. 3), where $E$ is
energy per nucleon. We will use this cross-section to calculate the
$\gamma$-ray luminosity below.

Next, consider the shocks embedded in the stellar winds
of massive stars. Owocki, Castor and Rybicki (1988) have shown
that these shocks can have velocities ($U_s$) as
large as the wind velocity
($U_w$) in the observer's frame. For massive O or B stars, the wind
velocities are observed to be close to $0.01c$ and mass losses
are of the order of $10^{-5}$ M$_{\odot}$ per yr (Bieging, Abbott and
Churchwell 1989). Suppose that a fraction of $\eta _p$ of the kinetic
energy
in the shock is deposited in the cosmic ray particles
($\eta _p \sim 0.1$ in, e.g., Blandford and Eichler 1987).
The spectrum of cosmic ray particles
at high energies, where $E \sim pc$, is observed to be
a power law in energy $E$, but theoretically, the spectrum
is expected to be a power law in momentum $p$ (e.g., Drury 1983).
We will use a spectral index $\alpha$ of $2.3$ in momentum in
our calculation (Biermann and Cassinelli 1993). One can then
write the kinetic energy of cosmic rays with
 the number density
$N(E)d(E) \propto E^{- \alpha} dE$
observed at relativistic energies , as

$$
\eta _p \rho _w U_s ^2 = \int A ({pc \over \GeV})^{- \alpha} E_k
d({pc \over \GeV}) \>, \eqno(1)
$$

where $\rho _w$ is the density in the stellar wind, $A$ (cm$^{-3}$)
denotes the abundance of cosmic ray protons and $E_k$
is the kinetic energy.

At energies $\sim$ GeV per nucleon, the abundance of oxygen nuclei
in cosmic rays is observed to be $\sim 3.5 \times
10^{-3}$ with respect to that of the protons. In other words,
the spectrum for oxygen nuclei, $N_O (pc) d(pc)$ can be written as
$N_O (pc) d(pc) = A_O ({pc \over 16 \GeV})^{- \alpha} d({pc \over 16
\GeV})$,
where $A_O=3.5 \times 10^{-3} A$. However, stellar winds
of OB stars are not enriched and the abundance
of heavy nuclei will be smaller than the standard abundance
in the cosmic ray
particles. If we we use the abundances of oxygen nuclei in the
solar system (denoting it by $A_{O,s}=6.7 \times 10^{-4} A$) then
we will obtain a lower limit of the flux. We note here that
first ionization potential effects can increase the abundance
of elements of low ionization potential, so that $A_O$ may be
somewhat larger than $A_{O,s}$.

The luminosity per proton (erg/s) can then be expressed as
$ E_\gamma \int N_O (pc) \> \sigma (pc) \> v \> d(pc)$,
where $E_\gamma$ is the energy of the emitted photon
(here, $4.4$ MeV). The
total luminosity of the $4.4$ MeV line from a region of with
volume $V$ and density $\rho$, is given by

$$
L_{\gamma} = n _a V  \>
E_\gamma \> \Bigl ( \int N_O (pc)
\sigma (pc) \> v \> d(pc) \Bigr ) \>. \eqno(2)
$$

Here, $n _a$ denotes the density of proton in the ambient
dense gas. For a typical value of the density, we can use the
numbers that Genzel and Stutzki (1989) have tabulated for the
physical parameters of the dense molecular gas surrounding
the OB stars in Orion A (their Table 1). The typical density is
$n_a \sim 10^5$ cm$^{-3}$ and kinetic temperature, $T \sim 100$ K.

The mass density in the stellar wind, $\rho _w$, in eqn $(1)$
can be written as
$
\rho _w = (\dot M ^2 / 4 \pi r^2 U_w) \>,
$
where $\dot M$ is the mass loss of the star, $r$ is the distance from
the star and $U_w$ is the
stellar wind velocity. Here the relevant volume is $ \sim \pi r^3$,
where $r$ is the current radius of the shock in the
stellar wind. Consider a completely spherical shock travelling
outwards through the stellar wind. The time evolution of the
$\gamma$-ray line luminosity is thus an initial rise, when particle
acceleration starts; a $1/t$ decay follows with a strong rise
to a maximum when the shock hits the surrounding material.
The radius $r$ where we envisage most of the the production of
$\gamma$-ray photons, that is, where the shocks meet the
surrounding dense molecular or atomic gas, is where the ram
pressure of the stellar wind balances the thermal
pressure of the gas. In other words, $(\dot M /4 \pi r^2 U_w) U_w ^2
\sim n_a k T$, which leads to,

$$
r_{pc} \sim 0.62 \> n_{a,5}^{-1/2}  T_2 ^{-1/2} (\dot M_{-5})^{1/2}
U_{w, -2.5}^{1/2} \>. \eqno(3)
$$

Here, $r$ is in parsecs, $n_a=10^5 n_{a,5}$ cm$^{-3}$, $T=100
\> T_2$ K,
$\dot M=10^{-5} \dot M_{-5}$ M$_{\odot}$ per yr, and $U_w=10^{-2.5}
U_{w, -2.5} \> c$.


Shock in the wind for OB or Wolf Rayet stars are only marginally
super-Alfv\'enic or even sub-Alfv\'enic at the equator. The relevant Alfv\'en
speed is proportional to $\sin \theta$ along a spherical surface such as a
shock in our simplified picture. The Alfv\'enic Mach number of the
shock, and the density jump, are very small near the equator, and the
spectrum is steep. Steeper spectra give rise to more $\gamma$-ray line
photons and a higher ratio of $\gamma$-ray line to the $\pi^o$-decay
luminosity. We have calculated both luminosities as a function of the
equatorial Alfv\'enic Mach number and find that for a Mach number that
is below unity at the equator, the $\gamma$-ray line luminosity is much
higher than in eqn ($4$), {\it and} the ratio of the line luminosity at
$4.4 \MeV$ to the $\pi^o$-decay luminosity at $100 \MeV$ is large enough to
explain the data. In the following we use the case of the equatorial
equivalent Alfv\'enic Machnumber 0.5, close to what we estimated on the basis
of the nonthermal radio emission (Biermann, Strom, Falcke 1994); one
consequence is that no particles are accelerated in an equatorial belt.

Combining equations $(1), (2),$ and $(3)$, and allowing for the latitude
variation of the Alfv\'enic Machnumber of the shock, one obtains the
$\gamma$-ray luminosity at $4.4$ MeV, as (using the cosmic abundance of oxygen
nuclei as a lower limit, as discussed above)

$$ \eqalignno{
L_\gamma &\ga 3 \times 10^{33} r_{pc} n_{a, 5} \> \eta _{p, -1}
\> (\dot M_{-5}) ( U_{s,-2.5} )^2 ( U_{w,-2.5})^{-1} \> {\rm erg/s} &\cr
&\approx 2 \times 10^{33} n_{a, 5}^{1/2} \>  T_2 ^{-1/2}
\> \eta _{p, -1} \> (\dot M_{-5})^{3/2} \> (U_{w,-2.5})^{-1/2} \>
( U_{s,-2.5} ) ^2 \> {\rm erg/s} \>,&(4)\cr}
$$

where $\eta _p= 0.1 \eta _{p,-1}$.
The opacity of the gas for these photons due to, for example,
Thomson scattering, is small for reasonable parameters of the
wind and the luminosity in the above expression is, therefore,
will not be diminished. At a mean distance of $\sim 0.45$ kpc
for the I Orion OB association
(Genzel and Stutzki 1989), and for $4.4$ MeV photons, this
means a flux of $5 \times 10^{-5}$ photons cm$^{-2}$
s$^{-1}$. The observed flux in the energy range $3-7$ MeV is
$\sim 10^{-4}$ photons cm$^{-2}$ s$^{-1}$, of which about
$50\%$ is thought to be due to the $4.4$ MeV line (Bloemen
\etal 1994).

There are $56$ O6 through B2 stars in the I Orion OB association
(Genzel and Stutzki 1989); as they also mention, there is evidence
mainly from infrared data that there may be a few more luminous
stars with strong winds, embedded in denser gas.
Therefore, the observed flux can be accommodated within the
framework of the above scenario, considering the uncertainty
in the geometry of the shocks in the wind, the actual
mass loss rate and the shock and wind velocities. We implicitly
assume that the duty cycle of shocks hitting the surrounding
medium is fairly high; the time-scale involved is of order
$10^3$ yrs. It is worth noting that, we do not require an increased
cosmic ray intensity in the clouds; we use only the parameters for
energetic particle acceleration in stellar winds, as supported by
radio observations of OB stars and radio supernovae (Biermann
and Cassinelli 1993).


With the cosmic abundance of oxygen nuclei in the ISM, it
is easy to see that the contribution to the above luminosity
from the collisions of energetic protons in cosmic rays with
ambient oxygen nuclei is same as in eqn ($4$). However, the
width of such a line will be much smaller due to the small
recoil velocity of the heavy nuclei in a collision with
energetic cosmic ray protons.

\bigskip

\centerline{3. DISCUSSION}
\medskip
Bloemen \etal (1994) noted the absence of any contribution
to the luminosity in the $3-7$ MeV range from $\pi ^o$ decay.
It is, therefore, worth while to calculate the $\gamma$-ray
luminosity due to decaying $\pi ^o$ from proton-proton collisions
in the above scenario. Using the cross-section for neutral
pion production in $p-p$ collisions given in Stephens
and Badhwar (1981), we find that the luminosity
in the pion-decay $\gamma$-ray photons at 100 MeV photon energy is,

$$
L_{\gamma , \pi ^o} \approx 3 \times 10^{33}
r_{pc} n_{a, 5} \> \eta _{p, -1}
\> (\dot M_{-5}) ( U_{s,-2.5} )^2 ( U_{w,-2.5})^{-1} \> {\rm erg/s}\>.
\eqno(5)
$$

Bloemen \etal (1994) have discussed the emissivity of the Orion
complex for energies more than $100 \MeV$, as was observed by
COS-B (Bloemen \etal 1984). The luminosity of the cloud complex
at $\sim 100 \MeV$ is $\sim (4-7) \times 10^{33}$ erg s$^{-1}$. This is
consistent with our derived expected fluxes.

We have also calculated the luminosity from bremsstrahlung
emitted by energetic electrons accelerated in the shocks.
Using the detected flux in the lines as the upper limit of
any background continuum, the electron bremsstrahlung does
not, however, lead to any additional constraints
on the ratio of electrons to protons in cosmic rays.

It is important to consider the energy loss of the accelerated
heavy nuclei in the stellar wind, apart from the line radiations
discussed above. The process of acceleration of the heavy
nuclei competes with energy losses due to interaction with
the plasma and it is important that they are indeed accelerated
to the required energies to be able to produce the observed $\gamma$-ray
lines. In the denser region near the stars, the energy loss
is large and one expects a critical length scale (the distance
from the star) only beyond which the injection of the heavy nuclei to
high energies is possible.

To estimate the critical length scale, we consider the energy
loss of the heavy nuclei in interactions with the thermal
plasma. At the postshock temperature of $T_{sh} = 1.4 \times 10^7
(U_s/ 10^3 {\rm km} {\rm s}^{-1})^2$ K, the gas will be highly ionized
(and therefore, ionization losses will not dominate). The
energy loss for low energy oxygen nuclei can be written as
(per acceleration cycle as in Biermann (1993)),

$$
\Delta \beta \approx 9.7 \times 10^{-6}
(U_{s,-2.5} )^{-3} (U_{w,-2.5} )^{-1}
\dot M_{-5} \> r_{pc}^{-1} \>, \eqno(8)
$$

If we
equate this loss with $U_s/c$, the gain per cycle in first order Fermi
acceleration, we obtain a critical length scale of
$r_{crit} \sim  10^{16} (U_{s,-2.5} )^{-4} (U_{w,-2.5} )^{-1}
\dot M_{-5}$ cm. In a supernova event the initial peak line luminosity
thus scales with the fifth power of the shock velocity
provided (a) the photons can escape, and (b) spallation does
not stop the acceleration altogether. This luminosity arises from the
shock travelling through the stellar wind.  Thus, the nuclei can readily be
accelerated as described in our scenario above.

There is possibly a large contribution to accelerated particles
from the steady stand-off
shock of the stellar wind when it encounters the dense circumstellar
material. If cosmic rays were injected at that shock, and cosmic
rays would influence the shock structure, then a strong instability
develops (Zank \etal 1990). In that case, the dense circumstellar
gas is likely to be mixed into the shock region, making injection
of energetic particles rather difficult. Therefore we emphasize
that the shock travelling through the stellar wind and then
hitting the outer boundary of the wind zone is more likely
to be effective in accelerating particles.

The energetic particles from these shocks ionize only a small fraction of the
overall cloud volume, so that outside the shells a cosmic ray spectrum with a
turnover, or a cutoff, at low energies will be expected. The cutoff/turnover
for
cosmic rays in the ISM is expected to be $\sim 30 \MeV$ for protons from the
observations of ionization rate in the diffuse clouds in the ISM (Nath and
Biermann 1994).  The ionization loss from interaction of the energetic
particles in the wind shell limits the the grammage seen to a few g/cm$^2$,
otherwise the ionization loss would eliminate those particles which are
responsible for the excitation of the energetic C and O atoms; beyond the
shells low energy protons arising from the wind shocks are no longer effective.


The production of $^{26}$Al in the scenario proposed may be relevant
in accounting for its observed high abundance (Lee \etal 1977;
Clayton 1994); we will consider this question in a separate
communication.

There are various theories of cosmic ray origin that may explain
the $\gamma$-ray line observations; in particular, the works
of Bykov \etal (Bykov and Toptygin 1990; Bykov and Toptygin 1992;
Bykov and Fleishman 1992; Bykov and Bloemen 1994), who argue for
acceleration by a collection of shock waves.
In the theory of cosmic ray origin proposed by Biermann \etal
(Biermann 1993; Biermann and Cassinelli 1993; Stanev \etal 1993;
 Rachen \etal 1993), one important
argument is the acceleration of cosmic ray nuclei in SN-shocks
that go through stellar winds. These shocks are argued to provide
particle energies to $\sim 3 \times 10^9 \GeV$, dominated by
heavy elements. The interpretation given here is thus an important
consistency check on this proposal.

\bigskip

\centerline{CONCLUSIONS}\medskip

We have shown that the recently detected $\gamma$-ray lines
in $3-7$ MeV range by COMPTEL in the Orion complex can be
explained in terms of emissions from energetic heavy nuclei
accelerated in the shocks of stellar winds of massive stars
as the nuclei collide with the dense and cool gas in the
surroundings.
Our calculations does not contain any free parameter and the
luminosity is derived as a function of the ambient density,
mass loss rate of massive stars, velocities of the stellar w
inds and the shocks within.  We emphasize that the stellar parameters used are
the same as those which we need to interpret the nonthermal radio emission.
stellar winds.

We thus provide here a consistency check on the theory that nuclei get
accelerated in shocks in the winds of massive stars.

\bigskip
\noindent
{\bf Acknowledgments}\medskip
We wish to thank Dr. R. Diehl for extensive discussion
on the $\gamma$-ray data, and Dr. H. Bloemen for pointing out the
relevance of the COS-B observations at high energies. BBN thanks
the Max Planck Society for a fellowship.

\bigskip\bigskip\bigskip
 \vfill \eject
\centerline{REFERENCES}\medskip

\r{Bieging, J. H., Abbott, D. C., \& Churchwell, E. B. 1989, ApJ,
340, 518} \par
\r{Biermann, P. L. 1993, A\&A, 271, 649} \par
\r{Biermann, P. L., \& Cassinelli, J. P. 1993, A \& A, 271, 691} \par
\r{Biermann, P.L., Strom, R.G., Falcke, H. 1994 A \& A, submitted} \par
\r{Blandford, R. D., \& Eichler, D. 1987, Physics Reports, 154, 2} \par
\r{Bloemen, H. \etal 1984 A \& A, 139, 37} \par
\r{Bloemen, H. \etal 1994 A \& A, 281, L5} \par
\r{Bykov, A., \& Toptygin, I. N. 1990, Sov. Phys. JETP, 71, 702} \par
\r{Bykov, A., \& Toptygin, I. N. 1992, Sov. Phys. JETP, 74, 462} \par
\r{Bykov, A., \& Fleishman, G. D. 1992, MNRAS, 255, 269} \par
\r{Bykov, A., \& Bloemen, H. 1994, A \& A, 283, L1} \par
\r{Cameron, A. G. W. 1994, Nature, 368, 192} \par
\r{Clayton, D. D. 1994, Nature, 368, 222} \par
\r{Drury, L.O'C 1983, Rep. Prog. in Phys., 46, 973} \par
\r{Genzel, R., \& Stutzki, J. 1989, ARAA, 27, 41} \par
\r{Lee, T., Papanastassiou, D. A., \& Wasserburg, G. J. 1977, ApJ,
211, L107} \par
\r{MacFarlane, J. J., \& Cassinelli, J. P. 1989, ApJ, 347, 1090} \par
\r{Meneguzzi, M., \& Reeves, H. 1975, A \& A, 40, 91} \par
\r{Nath, B. B., \& Biermann, P. L. 1994, MNRAS, 267, 447} \par
\r{Owocki, S. P., Castor, J. I., \& Rybicki, G. B. 1988, ApJ, 235, 914}
 \par
\r{Rachen, J. P. \etal 1993, A\&A, 273, 377} \par
\r{Ramaty, R., Kozlovsky, B., \& Lingenfelter, R. 1979 ApJS, 40, 487}
\par
\r{Stanev, T. \etal 1993, A\&A, 274, 902} \par
\r{Stephens, S. A., \& Badhwar, G. D. 1981, Astrophys. Sp. Sc., 76, 213} \par
\r{White, R. L. 1985, ApJ, 289, 698} \par
\r{White, R. L., \& Chen, W. 1992, ApJ, 387, L81} \par
\r{Zank, G. P. 1990, A\&A, 233, 275} \par

\vfill \eject
\end